\documentclass[12pt]{article}
\usepackage{amsmath}
\usepackage{amssymb}
\usepackage[mathscr]{eucal}
\usepackage{authblk}
\usepackage{cite}
\usepackage{pdfsync}
\usepackage[
bookmarks=true,
backref=true,
colorlinks=true,
linkcolor=red,
citecolor=blue,
urlcolor=green]{hyperref}

\newcommand{\gen}[1]{\partial_{#1}}

\newtheorem{lemma}{Lemma}
\newtheorem{prop}{Proposition}
\newtheorem{thm}{Theorem}
\newtheorem{cor}{Corollary}

\numberwithin{equation}{section}
\numberwithin{thm}{section}
\numberwithin{lemma}{section}
\numberwithin{prop}{section}
\numberwithin{cor}{section}
\numberwithin{rmk}{section}
\numberwithin{defn}{section}
\DeclareMathOperator{\Sl}{sl}

\DeclareMathOperator{\So}{so}

\DeclareMathOperator{\diag}{diag}
\newcommand{\Conf}{\mathsf{C}}
\newcommand{\conf}{\mathsf{AC}}

\newcommand{\simil}{\mathsf{A\tilde{P}}}
\DeclareMathOperator{\Ort}{O}
\DeclareMathOperator{\ort}{o}
\DeclareMathOperator{\poinc}{p}
\DeclareMathOperator{\euc}{e}

\newtheorem{example}{Example}

\begin{document}

\title{Lie symmetry structure of nonlinear wave equations in $(n+1)$-dimensional space-time}

\author[1]{P.~Basarab-Horwath\thanks{e-mail: bhrwthp@gmail.com}}
\author[2]{F.~G\"ung\"or\thanks{e-mail: gungorf@itu.edu.tr}}
\author[2]{C.~\"Ozemir\thanks{e-mail: ozemir@itu.edu.tr}}

\affil[1]{Nygatan 44, 582 27 Link\"oping, Sweden}
\affil[2]{Department of Mathematics, Faculty of Science and Letters, Istanbul Technical University, 34469 Istanbul, Turkey}

\date{}

\maketitle

\abstract{We study Lie point symmetry structure of generalized nonlinear wave equations of the form $\Box u=F(x, u, \nabla u)$ where $\Box$ is the $(n+1)$-dimensional space-time wave (or d'Alembert) operator, $x\in \mathbb{R}^{n+1}$ ($n\geq 2$).}


\section{Introduction}

The aim of this paper is to study group-theoretical properties of the class of generalized wave equations of the form
\begin{equation}\label{waveqn1}
\Box u=F(x, u, \nabla u), \quad n\geq 2
\end{equation}
and its subclass consisting of the equations of the form
\begin{equation}\label{waveqn2}
\Box u=F(x, u),  \quad n\geq 2,
\end{equation}
where $\displaystyle \Box u=g^{\mu\nu}u_{\mu\nu}$ and $\nabla u=(u_0, u_1,\dots, u_n)$, $x=(x^0, x^1, x^2,  \ldots, x^n)$. Here, $\displaystyle u_{\mu}=\partial u/\partial x^{\mu},\; u_{\mu\nu}=\partial^2 u/\partial x^{\mu}\partial x^{\nu}$ for $\mu, \nu= 0, 1, \dots, n$ and $\displaystyle g^{\mu\nu}$ denotes the $(n+1)\times(n+1)$ matrix given by $\displaystyle g^{\mu\nu}=\diag(1,-1,\dots, -1)$. The class \eqref{waveqn2} includes physically and mathematically significant special cases  like, among others, the  linear  wave equation
\begin{equation}\label{lin-wave}
  \Box u=0,
\end{equation}
the Klein--Gordon equation with a potential $V(x)$
\begin{equation}\label{KG}
  \Box u+V(x)u=0,
\end{equation}
the semilinear Klein--Gordon equation
\begin{equation}\label{semi-KG}
  \Box u+m^2 u+\gamma u^p=0,   \quad p\ne 1,
\end{equation}
which plays a central role in  quantum field theory, the sine-Gordon equation
\begin{equation}\label{sin-G}
 \Box u+\sin u=0,
\end{equation}
which also arises in quantum field theory, but was first studied in differential geometry
and Yamabe equation again in the context of differential geometry. Conformal invariance of \eqref{lin-wave} in $(1+3)$-dimension has long been known  \cite{Bateman1909, Cunningham1910, Carmichael1927}. Here we  concentrate on establishing invariance properties of the extensions of the linear wave equation involving arbitrary dependence on temporal-spatial variables, wave function and its first-order derivatives. We make use  of the notion of equivalence group of the equation to determine possible canonical forms.

This paper is organized as follows. In Section 2 we recall some necessary definitions. In section 3 we calculate equivalence transformations of equations of the form \eqref{waveqn1}. In this paper, an equivalence transformation is an invertible point-transformation (dependent only on the arguments $(x,u)$) which preserves the form of \eqref{waveqn1}, and in particular transforms the non-linearity $F(x,u,\nabla u)$ into a non-linearity of the same type. These are given in Propositions \ref{equivalence1} and \ref{equivalence2}.  The symmetry condition for equation \eqref{waveqn1} can then be obtained by requiring that an equivalence transformation preserve the given non-linearity; this is given in Section 4. Sections   5 an 6 are devoted to identify some canonical forms based on the symmetry structure determined in the previous two sections.

\section{Notation and definitions}
In this section, in order to make this paper self-contained  we present some definitions that will be used throughout.
The $(n+1)$-dimensional Minkowski space $\mathbb{M}^{1,n}$ is the pseudo-Euclidean space $\mathbb{R}^{1,n}$ which is equipped with a Lorentz scalar product (the Minkowski metric) defined by
\begin{equation}\label{scalar-prod}
  x\cdot y=g_{\mu\nu}x^{\mu}y^{\nu}=x^0y^0-x^1y^1-\cdots-x^ny^n.
\end{equation}
A non-zero function $\gamma(x)=x\cdot x\in \mathbb{M}^{1,n}$ is called timelike if $\gamma(x)>0$, light-like if $\gamma(x)=0$ and spacelike if $\gamma(x)<0$. The zero vector $x=0$ is considered spacelike. The set of light-like vectors is called the light-cone of $\mathbb{M}^{1,n}$.

A conformal transformation is one which preserves the metric \eqref{scalar-prod} up to a scalar factor. A vector field $X$ is an infinitesimal conformal transformation or conformal Killing field if its local flow  $\phi_t$ (a local one-parameter group) consists of conformal transformations for all $t$ in some neighbourhood of zero. Thus, if $X$ to be a conformal Killing  field we have the relation
\begin{equation}\label{conf-killing}
  \phi_t^{*}g=\lambda_t g
\end{equation}
for some positive function $\lambda_t$, or
using the definition
$$(\phi_t^{*}g)(x)=g(x)+t\mathcal{L}_Xg(x)+\mathcal{O}(t^2)$$
in terms of the Lie derivative
\begin{equation}\label{killing-cond}
  \mathcal{L}_Xg=\frac{d}{dt}(\phi_t^{*}g)(x)\big\vert_{t=0}=\kappa g
\end{equation}
for some not necessarily positive function $\kappa(x)$.
Here
$$g(x)=g_{\mu\nu}dx^{\mu}\otimes  dx^{\nu}$$
is the Lorentzian metric tensor field.
Condition \eqref{killing-cond} is known as Killing's equation. A solution $X$ of the Killing's equation is called Killing field. It can be shown that the Killing equation  \eqref{killing-cond} is equivalent to
\begin{equation}\label{killing-cond-2}
  g_{\alpha\rho}\partial_{\beta}\xi^{\rho}+g_{\beta\sigma}\partial_{\alpha}\xi^{\sigma}=\kappa g_{\alpha\beta},
\end{equation}
where $\xi^{\mu}$ are the coefficients of the Killing field $X$.
The reader is referred to the books \cite{CrampinPirani1987, Schottenloher2008} for a discussion of these concepts.

A conformal transformation for which the Killing factor $\kappa$ is a constant is called a homothety. If $\kappa=0$, the Killing field $X$ generates infinitesimal isometries (group of motions) of the metric and these isometries form a Lie algebra.

The conformal Lie algebra $\conf(1,n)$  is a $(n+2)(n+3)/2$-dimensional algebra with the basis elements $P_{\mu}$, $J_{\mu\nu}$, $D$  and $K_{\mu}$, $\mu=0,1,2,\ldots,n$.
They satisfy the commutation relations\cite{BarannykBasarabHorwathFushchych1998}
\begin{equation}\label{comm}
  \begin{split}
      & [P_{\mu},P_{\nu}]=0, \quad  [P_{\lambda},J_{\mu\nu}]=g_{\lambda\mu}P_{\nu}-g_{\lambda\nu}P_{\mu},\\\
      & [J_{\mu\nu},J_{\rho\sigma}]= g_{\mu\sigma} J_{\nu\rho} +  g_{\nu\rho} J_{\mu\sigma}-
      g_{\mu\rho}J_{\nu\sigma} - g_{\nu\sigma}J_{\mu\rho}J_{\nu\sigma},\\\
      & [P_{\mu},D]=P_{\mu},  \quad [D,J_{\mu\nu}]=0,  \quad [D,K_{\mu}]=-K_{\mu},\\\
      & [K_{\lambda},J_{\mu\nu}]=g_{\lambda\mu}K_{\nu}-g_{\lambda\nu}K_{\mu},  \quad [K_{\lambda},K_{\mu}]=0,  \quad [P_{\mu},K_{\nu}]=2(g_{\mu\nu}D-J_{\mu\nu}).
  \end{split}
\end{equation}
The basis elements of the algebra generate translations, rotations and Lorentz boosts, dilations and special conformal transformations, respectively.   The subalgebras spanned by $\{P_{\mu}, J_{\mu\nu}\}$ and  $\{P_{\mu}, J_{\mu\nu}, D\}$ are called the  Poincar\'e algebra $\poinc(1,n)$  of dimension $(n+1)(n+2)/2$ and the similitude algebra $\simil(1,n)$ of dimension $(n^2+3n+4)/2$. The similitude algebra has the structure of semidirect sums of the form
\begin{equation}\label{struct}
  \simil(1,n)=\{D\}\oplus_s\{\ort(1,n)\oplus_s \mathbb{R}^{n+1}\}.
\end{equation}
The conformal group $\mathsf{C}(1,n)$ of the Minkowski space $\mathbb{M}^{1,n}$ is locally isomorphic to the pseudo-orthogonal group $\Ort(n+1,2)$.

\section{Equivalence transformations} In this section we calculate the equivalence groups of the non-linear wave equations \eqref{waveqn1} and \eqref{waveqn2}.

We look at transformations $\tilde{x}^{\mu}=X^{\mu}(x, u),\: \tilde{u}=U(x,u)$ which preserve the form of the equation (\ref{waveqn1}). Thus we must obtain another equation of the form

\[\label{transformedwaveqn1}
\Box \tilde{u}=\tilde{F}(\tilde{x}, \tilde{u},  \nabla{\tilde{u}}).
\]
These transformations must be invertible, so we require that $dX^0\wedge dX^1\wedge\dots dX^n\wedge dU\neq 0$.

First, note that
\begin{equation}
 D_{\mu}U=D_{\mu}\tilde{u}
=\tilde{u}_{\tilde{\sigma}}D_{\mu}\tilde{x}^{\sigma}
=\tilde{u}_{\tilde{\sigma}}D_{\mu}X^{\sigma},
\end{equation}
where $\displaystyle \tilde{u}_{\tilde{\sigma}}=\partial \tilde{u}/\partial \tilde{x}^{\sigma}$, and
$$D_{\mu}=\partial_{\mu}+u_{\mu}\partial_{u}, \quad \mu=0,1,2,\ldots,n,$$ is the truncated total derivative operator
and we sum over repeated indices. Thus we have
\[
U_{\mu}+u_{\mu}U_u=\tilde{u}_{\tilde{\sigma}}(X^{\sigma}_{\mu}+u_{\mu}X^{\sigma}_u),
\]
from which we have
\[
(U_u-X^{\sigma}_u\tilde{u}_{\tilde{\sigma}})u_{\mu}=\tilde{u}_{\tilde{\sigma}}X^{\sigma}_{\mu}-U_{\mu}.
\]
Further, we have
\begin{equation}\label{}
  D_{\mu}D_{\nu}U=D_{\mu}(\tilde{u}_{\tilde{\sigma}}D_{\nu}X^{\sigma})
=\tilde{u}_{\tilde{\sigma}\tilde{\rho}}D_{\mu}X^{\rho}D_{\nu}X^{\sigma}+ \tilde{u}_{\tilde{\sigma}}D_{\mu\nu}X^{\sigma}.
\end{equation}
We have also
$$  D_{\mu} D_{\nu}U=U_{\mu\nu} + u_{\mu}U_{\nu u} + u_{\nu}U_{\mu u} + u_{\mu}u_{\nu}U_{uu} + u_{\mu\nu}U_u$$
and a similar expression for $D_{\mu} D_{\nu}X^{\sigma}$. Solving for $u_{\mu\nu}$ we obtain
\begin{equation}\label{definingmunu}
[U_u-\tilde{u}_{\tilde{\sigma}}X^{\sigma}_u]u_{\mu\nu}=\tilde{u}_{\tilde{\sigma}\tilde{\rho}}D_{\mu}X^{\rho}D_{\nu}X^{\sigma} + \tilde{u}_{\tilde{\sigma}}V_{\mu\nu}\cdot X^{\sigma}-V_{\mu\nu}\cdot U,
\end{equation}
where
$$ V_{\mu\nu}\cdot U=U_{\mu\nu} + u_{\mu}U_{\nu u} + u_{\nu}U_{\mu u} + u_{\mu}u_{\nu}U_{uu}$$ and a similar expression for $V_{\mu\nu}\cdot X^{\sigma}$. With $\Box u=g^{\mu\nu}u_{\mu\nu}$ we obtain from (\ref{definingmunu})

\begin{equation}\label{transformedeqn}
[U_u-\tilde{u}_{\tilde{\sigma}}X^{\sigma}_u]\Box u= \tilde{u}_{\tilde{\sigma}\tilde{\rho}}g^{\mu\nu}D_{\mu}X^{\rho}D_{\nu}X^{\sigma} + \tilde{u}_{\tilde{\sigma}}V\cdot X^{\sigma}-V\cdot U
\end{equation}
with
$$V\cdot U=\Box U + 2g^{\mu\nu}u_{\mu}U_{\nu u} + g^{\mu\nu}u_{\mu}u_{\nu}U_{uu}$$ and a similar expression for $V\cdot X^{\sigma}$.

In (\ref{transformedeqn}) we must have no mixed derivatives $\tilde{u}_{\tilde{\sigma}\tilde{\rho}}$ for $\rho\neq \sigma$. Also, the remaining terms in $\displaystyle \tilde{u}_{\tilde{\sigma}\tilde{\rho}}g^{\mu\nu}(D_{\mu}X^{\rho})D_{\nu}X^{\sigma}$ must give us $\lambda \Box \tilde{u}$ for some non-zero function $\lambda$. From these requirements, we find that we have
\begin{equation}\label{conds}
  \begin{split}
 g^{\mu\nu}D_{\mu}X^0D_{\nu}X^0&=\lambda, \\
 g^{\mu\nu}D_{\mu}X^aD_{\nu}X^a&=-\lambda,\quad a=1,\dots, n,\\
 g^{\mu\nu}D_{\mu}X^{\rho}D_{\nu}X^{\sigma}&=0,\quad \rho\neq \sigma.
  \end{split}
\end{equation}
Note that we must have $\lambda >0$. If not, then the vectors $(D_{0}X^a, D_{1}X^a,\dots, D_{n}X^a)$ for $a=1,\dots, n$ are timelike and pairwise pseudo-orthogonal, that is, we have $$(D_{0}X^a)D_{0}X^b-(D_{1}X^a)D_{1}X^b-\dots - (D_{n}X^a)D_{n}X^b=0$$
for $a\neq b$. This is impossible, for if two vectors $(x_0, \mathbf{x}),\; (y_0,\mathbf{y})$ are pseudo-orthogonal, we have $x_0y_0-\mathbf{x}\cdot\mathbf{y}=0$ and by the Cauchy--Schwarz inequality we have $x_0^2y_0^2\leq (\mathbf{x}\cdot\mathbf{x})(\mathbf{y}\cdot\mathbf{y})$ and then we have $(x_0^2-\mathbf{x}\cdot\mathbf{x})y_0^2\leq (\mathbf{x}\cdot\mathbf{x})(\mathbf{y}\cdot\mathbf{y}-y_0^2)$ and it is impossible to have the two inequalities $x_0^2-\mathbf{x}\cdot\mathbf{x}>0,\; y_0^2-\mathbf{y}\cdot\mathbf{y}>0$ simultaneously. Thus, the vector $(D_{0}X^0, D_{1}X^0,\dots, D_{n}X^0)$ is timelike and $(D_{0}X^a, D_{1}X^a,\dots, D_{n}X^a)$ is spacelike for $a=1,\dots, n$.

From
\[
g^{\mu\nu}(D_{\mu}X^{\rho})D_{\nu}X^{\sigma}=0,\quad \rho\neq \sigma
\]
we have, on differentiating with respect to $u_{\alpha}$

\[
g^{\alpha\nu}X^{\rho}_u(X^{\sigma}_{\nu} + u_{\nu}X^{\sigma}_u)+ g^{\mu\alpha}X^{\sigma}_u(X^{\rho}_{\mu} + u_{\mu}X^{\rho}_u)=0,\;\; \rho\neq \sigma.
\]
Since $g^{\alpha\beta}=0,\;\; \alpha\neq\beta$ and $g^{\alpha\beta}=\pm 1$ when $\alpha=\beta$ we obtain

\[
X^{\rho}_u(X^{\sigma}_{\alpha} + u_{\alpha}X^{\sigma}_u)+X^{\sigma}_u(X^{\rho}_{\alpha} + u_{\alpha}X^{\rho}_u)=0.
\]
From this we see that
\[
X^{\rho}_uX^{\sigma}_u=0,\quad X^{\rho}_uX^{\sigma}_{\alpha}+ X^{\sigma}_uX^{\rho}_{\alpha}=0,\;\; \rho\neq \sigma.
\]
If $X^{\rho}_u=0,\; X^{\sigma}_u\neq 0$ we obtain $X^{\rho}_{\alpha}=0$ which then gives us $dX^{\rho}=0$ which contradicts the requirement that $dX^0\wedge dX^1\wedge\dots\wedge dX^n\wedge dU\neq 0$. Hence we must have $X^{\rho}_u=0$ for all $\rho= 0,1, \dots, n$. Since we have $U_u-X^{\sigma}_u\tilde{u}_{\tilde{\sigma}}\neq 0$ we then have $U_u\neq 0$, and now follows that $D_{\mu}X^{\rho}=\partial_{\mu}X^{\rho}=X^{\rho}_{\mu}$ and we can then rewrite our conditions on the $X^{\rho}$ as
\[
g^{\mu\nu}X^{\rho}_{\mu}X^{\sigma}_{\nu}=\lambda(x)g^{\rho\sigma},
\]
where $\lambda(x)>0$. This equation then shows that the transformations $\Phi(x)=(X^0, X^1,\dots, X^n)\in \Conf(1,n)$, the conformal group in $(n + 1)$-space-time.

Returning to equation (\ref{transformedeqn}), we now have
\[
U_u\Box u=\lambda(x)\Box\tilde{u} + \tilde{u}_{\tilde{\sigma}}\Box X^{\sigma} - (\Box U + 2g^{\mu\nu}u_{\mu}U_{\nu u} + g^{\mu\nu}u_{\mu}u_{\nu}U_{uu}).
\]
On putting $\Box u=F(x,u,\nabla u)$ and $\Box \tilde{u}=\tilde{F}(\tilde{x}, \tilde{u}, \nabla{\tilde{u}})$, we then have
\begin{equation}\label{reducedtransformed}
U_uF(x,u,\nabla u)=\lambda(x) \tilde{F}(\tilde{x}, \tilde{u}, \nabla{\tilde{u}}) + \tilde{u}_{\tilde{\sigma}}\Box X^{\sigma} - (\Box U + 2g^{\mu\nu}u_{\mu}U_{\nu u} + g^{\mu\nu}u_{\mu}u_{\nu}U_{uu}).
\end{equation}
This gives the relation between $F$ and the transformed non-linearity $\tilde{F}$. Thus our equivalence group consists of transformations $\tilde{x}^{\mu}=X^{\mu}(x), \;\; \tilde{u}=U(x, u)$ with $(X^0, X^1,\dots, X^n)\in \Conf(1,n),\; U_u\neq 0$, and any such transformation defines an equivalence transformation. We state this as the following result:
\begin{prop}\label{equivalence1} The equivalence group of Eq. \eqref{waveqn1}
is given by transformations
\begin{equation}\label{equiv-group-1}
\tilde{x}^{\mu}=X^{\mu}(x),\quad \tilde{u}=U(x,u), \quad \mu=0, 1, \dots, n,
\end{equation}
together with \eqref{reducedtransformed}, where $U_u\neq 0$ and the transformations $(X^0,X^1,\dots, X^n)\in \Conf(1,n)$, the conformal group in $(n+1)$-space-time, that is they satisfy
\[
g^{\mu\nu}\frac{\partial X^{\alpha}}{\partial x^{\mu}}\frac{\partial X^{\beta}}{\partial x^{\nu}}=\lambda(x)g^{\alpha\beta},
\]
with $g^{\mu\nu}=\diag (1, -1,\dots, -1)$ and $\lambda(x)>0$ is a smooth function given by
\[
\lambda(x)=g^{\mu\nu}X^0_{\mu}(x)X^0_{\nu}(x).
\]
\end{prop}
If in
\[
U_u\Box u=\lambda(x)\Box\tilde{u} + \tilde{u}_{\tilde{\sigma}}\Box X^{\sigma} - (\Box U + 2g^{\mu\nu}u_{\mu}U_{\nu u} + g^{\mu\nu}u_{\mu}u_{\nu}U_{uu})
\]
we replace $\Box u$ and $\Box\tilde{u}$ by $F(x,u)$ and $\tilde{F}(\tilde{x}, \tilde{u})$, we obtain
\begin{equation}\label{reducedtransformed1}
U_uF(x,u)=\lambda(x) \tilde{F}(\tilde{x}, \tilde{u}) + \tilde{u}_{\tilde{\sigma}}\Box X^{\sigma} - (\Box U + 2g^{\mu\nu}u_{\mu}U_{\nu u} + g^{\mu\nu}u_{\mu}u_{\nu}U_{uu}).
\end{equation}
Since the nonlinearities must be independent of the first-order derivatives, the terms linear and quadratic  in $\tilde{u}_{\tilde{\sigma}}$ must vanish. Consequently we have
\[
U_{uu}=0
\]
from the terms quadratic in $\tilde{u}_{\tilde{\sigma}}$. The terms linear in $\tilde{u}_{\tilde{\sigma}}$ give us, after substituting $u_{\mu}= ({\tilde{u}_{\tilde{\sigma}}X^{\sigma}_{\mu}}- {U_{\mu}})/{U_u}$,
\[
\Box X^{\sigma}=2g^{\mu\nu}\frac{X^{\sigma}_{\mu}U_{\nu u}}{U_u}
\]
and equation (\ref{reducedtransformed1}) simplifies to
\begin{equation}\label{reducedtransformed2}
U_uF(x,u)=\lambda(x) \tilde{F}(\tilde{x}, \tilde{u})  - \Box U + 2g^{\mu\nu}\frac{U_{\mu}U_{\nu u}}{U_u}.
\end{equation}

Clearly, we have $U(x,u)=A(x)u + B(x)$ with $A(x)\neq 0$ which is related to $X^{\sigma}$ through
\[
\Box X^{\sigma}=2g^{\mu\nu}\frac{X^{\sigma}_{\mu}A_{\nu}}{A}.
\]
So we have the following result:
\begin{prop}\label{equivalence2} The equivalence group of Eq. \eqref{waveqn2}
is given by transformations
\begin{equation}\label{eqiv-group-2}
\begin{split}
\tilde{x}^{\mu}&=X^{\mu}(x),\qquad \mu=0, 1, \dots, n,\\
\tilde{u}&=A(x)u + B(x),
\end{split}
\end{equation}
where the transformations $(X^0,X^1,\dots, X^n)\in \Conf(1,n)$, the conformal group in $(n+1)$-space-time, satisfying
\[
g^{\mu\nu}\frac{\partial X^{\alpha}}{\partial x^{\mu}}\frac{\partial X^{\beta}}{\partial x^{\nu}}=\lambda(x)g^{\alpha\beta},
\]
with $g^{\mu\nu}=\diag (1, -1,\dots, -1)$, $A(x)\neq 0$ is a smooth function satisfying
\[
\Box X^{\sigma}=2g^{\mu\nu}\frac{X^{\sigma}_{\mu}A_{\nu}}{A}
\]
and $\lambda(x)>0$ is a smooth function given by
\[
\lambda(x)=g^{\mu\nu}X^0_{\mu}(x)X^0_{\nu}(x).
\]
Under the transformation \eqref{eqiv-group-2}, $F$ and the transformed $\tilde{F}$ are related by
\begin{equation}\label{}
AF(x,u)=\lambda(x) \tilde{F}(\tilde{x}, \tilde{u})  - (\Box A)u-(\Box B)+ 2g^{\mu\nu}\frac{A_{\nu}}{A}(A_{\mu} u+B_{\mu}).
\end{equation}

\end{prop}

\section{The symmetry condition} In this section we calculate the condition for a vector field
$$Q=\xi^{\mu}(x,u)\partial_{\mu} +  \eta(x,u)\partial_u$$
to be the generator of a one-parameter point symmetry group of an equation of the form (\ref{waveqn1}).

By definition, a point symmetry of (\ref{waveqn1}) is a transformation which maps the set of solutions of (\ref{waveqn1}) onto itself, that is if $u(x)$ is a solution of (\ref{waveqn1}), then so is $\tilde{u}(\tilde{x})$; in other words, we have
\[
\Box \tilde{u}=F(\tilde{x}, \tilde{u}, \nabla{\tilde{u}}) \quad \text{when}\quad \Box u=F(x, u, \nabla u).
\]
Clearly, such a transformation is an equivalence transformation with $\tilde{F}=F$, and thus the vector field $Q$ is the generator of a (local) one-parameter group of equivalence transformations. Writing the one-parameter group of transformations as
\begin{align*}
\tilde{x}^{\mu}(x; \epsilon)&=X^{\mu}(x; \epsilon),\\
\tilde{u}(x,u; \epsilon)&=U(x,u; \epsilon),
\end{align*}
we see that the generators are
\[
\frac{d X^{\mu}}{d\epsilon}(x; \epsilon)\Bigr|_{\epsilon=0}=\xi^{\mu}(x),\quad \frac{d U}{d\epsilon}(x, u; \epsilon)\Bigr|_{\epsilon=0}=\eta(x,u).
\]
Differentiating equation (\ref{reducedtransformed}) with respect to $\epsilon$ at $\epsilon=0$ we obtain the symmetry condition
\[
\eta_uF=2\xi^0_0F + \xi^{\mu}F_{\mu} + \eta F_u + \tau_{\mu}F_{u_{\mu}} + u_{\sigma}\Box\xi^{\sigma} - \Box\eta - 2g^{\mu\nu}u_{\mu}\eta_{\nu u} - g^{\mu\nu}u_{\mu}u_{\nu}\eta_{uu},
\]
where $\tau_{\mu}=D_{\mu}\eta - u_{\sigma}D_{\mu}\xi^{\sigma}$ is the generator of the induced transformations of the derivatives $u_{\mu}$. The term $\tilde{u}_{\tilde{\sigma}}\Box X^{\sigma}$ gives, on differentiating with respect to $\epsilon$ at $\epsilon=0$, $\tau_{\sigma}\Box x^{\sigma} + u_{\sigma}\Box \xi^{\sigma}$, since
$$X^{\sigma}(x; \epsilon)=x^{\sigma} + \epsilon\xi^{\sigma} + \mathcal{O}(\epsilon^2),$$
and so $X^{\sigma}(x; \epsilon)|_{\epsilon=0}=x^{\sigma}$ and then $\Box x^{\sigma}=0$.  We have also used the fact that $$\lambda(x)=g^{\mu\nu}X^0_{\mu}(x; \epsilon)X^0_{\nu}(x; \epsilon)$$ when $\tilde{x}^{\mu}=X^{\mu}(x; \epsilon)$,  so that
\[
\kappa(x)=\frac{d\lambda}{d\epsilon}(x; \epsilon)\Bigr|_{\epsilon=0}=2\xi^0_0
\]
because
$$X^0(x; \epsilon)= x^0+ \epsilon\xi^0 + \mathcal{O}(\epsilon^2).$$
Further, the relations
\[
g^{\mu\nu}X^{\rho}_{\mu}X^{\sigma}_{\nu}=\lambda(x)g^{\rho\sigma}
\]
give, on differentiating with respect to $\epsilon$ at $\epsilon=0$,
\begin{equation}\label{killing-eqs}
  g^{\mu\sigma}\xi^{\rho}_{\mu} + g^{\nu\rho}\xi^{\sigma}_{\nu}=\kappa(x)g^{\rho\sigma}.
\end{equation}
These relations are precisely the Killing equations (see \eqref{killing-cond-2} for its geometrical definition) for the infinitesimal generators of the conformal Lie algebra $\mathsf{AC}(1,n)$. A nonzero nonconstant $\kappa(x)$ is called a conformal Killing factor.  The solution to this overdetermined system is called Killing solution and is given by (for example see \cite{FushchichShtelenSerov1993})
\begin{equation}\label{killing-sol}
  \xi^{\alpha}=2(c\cdot x)x^{\alpha}-c^{\alpha}x^2+b^{\alpha\beta}x_{\beta}
  +dx^{\alpha}+a^{\alpha},
\end{equation}
where $c\cdot x=c_{\beta}x^{\beta}$, $x^2=x_{\beta}x^{\beta}=g_{\alpha\beta}x^{\alpha}x^{\beta}$ and $c_{\beta}$, $b_{\alpha\beta}=-b_{\beta\alpha}$, $b_{0\beta}=b_{\beta 0}, b_{00}=b_{11}=\cdots =b_{nn}=0$, $d$, $a^{\alpha}$ $(\alpha,\beta=0,1,2,\ldots,n)$ are arbitrary integration constants. The part $\xi^{\alpha}_c=2(c\cdot x)x^{\alpha}-c^{\alpha}x^2$ of \eqref{killing-sol} is  conformal Killing solution with conformal factor  $\kappa_c(x)=4(c\cdot x)$. Tracing  Killing equations \eqref{killing-eqs} with $g_{\rho\sigma}$ we find the formula for the Killing factor
\begin{equation}\label{killing-factor}
  \kappa(x)=\frac{2}{n+1}\xi^{\mu}_{\mu}
\end{equation}
so that we can write $$\lambda(x;\varepsilon)=1+\frac{2\epsilon}{n+1}\xi^{\mu}_{\mu}+\mathcal{O}(\epsilon^2).$$
A general element of the conformal algebra has the form
\begin{equation}\label{gen-X}
  X=a^{\mu} P_{\mu}+b^{\mu\nu}J_{\mu\nu}+d D+c^{\mu}K_{\mu}
\end{equation}
and the defining relation \eqref{killing-cond} becomes
$$\mathcal{L}_Xg=\kappa(x)g=2(2c\cdot x+d) g.$$

This Lie algebra has dimension $\displaystyle \frac{(n+2)(n+3)}{2}$ and is spanned by the vector fields
\begin{equation}\label{killing-vf}
  \begin{split}
P_{\mu}&=\partial_{\mu},\\
J_{\mu\nu}&=x_{\mu}\partial_{\nu}-x_{\nu}\partial_{\mu},\\
D&=x^{\sigma}\partial_{\sigma},\\
K_{\mu}&=2x_{\mu}D-(x_{\alpha}x^{\alpha})\partial_{\mu},
\end{split}
\end{equation}
where $\mu, \nu= 0, 1, \dots, n$. We can summarise the above calculations in the following:
\begin{prop}\label{pro-4.1}
If $Q=\xi^{\mu}(x,u)\partial_{\mu} + \eta(x,u)\partial_u$ is the generator of a point-symmetry of equation (\ref{waveqn1}), then we have $\xi^{\mu}_u=0$ and the vector field $\xi^{\mu}(x)\partial_{\mu}$ is a linear combination of the $\displaystyle \frac{(n+2)(n+3)}{2}$ basis vector fields \eqref{killing-vf}.
Furthermore, the remaining part of the symmetry condition (classifying determining equation) is given by the equation
\begin{align}\label{symmcondition1}
& \xi^{\mu}F_{\mu} + \eta F_u + (\eta_{\mu} + u_{\mu}\eta_u-u_{\sigma}\xi^{\sigma}_{\mu})F_{u_{\mu}}+(2\xi^0_0-\eta_u)F \nonumber\\
& + u_{\sigma}\Box\xi^{\sigma} - \Box\eta - 2g^{\mu\nu}u_{\mu}\eta_{\nu u} - g^{\mu\nu}u_{\mu}u_{\nu}\eta_{uu}=0.
\end{align}
\end{prop}

The same calculation for the wave equation (\ref{waveqn2}), using equation (\ref{reducedtransformed2}), gives us the result:
\begin{prop}\label{pro-4.2}
If $Q=\xi^{\mu}(x,u)\partial_{\mu} + \eta(x,u)\partial_u$ is the generator of a point-symmetry of equation (\ref{waveqn2}), then we have $\xi^{\mu}_u=0$ and $\eta=a(x)u+b(x)$ and the vector field $\xi^{\mu}(x)\partial_{\mu}$ is a linear combination of the $\displaystyle \frac{(n+2)(n+3)}{2}$ basis vector fields \eqref{killing-vf}.
Furthermore, the remaining part of the symmetry condition (classifying determining equation) is given by the equation
\begin{equation}\label{symmcondition2}
  \xi^{\mu}F_{\mu} + \eta F_u +(2\xi^0_0-\eta_u)F - \Box\eta=0,\quad \Box\xi^{\mu}=2g^{\mu\nu}a_{\nu}(x).
\end{equation}
\end{prop}

We remark that the determining equations \eqref{symmcondition1}  and \eqref{symmcondition2} (first order linear PDEs in $F$) in  Propositions \ref{pro-4.1} and \ref{pro-4.2} permit us to solve for $\eta$ in terms of known Killing solutions $\xi^{\alpha}$ for a given $F$. Conversely, they may serve to classify all possible nonlinearities $F$ which are invariant under subalgebras of the full conformal algebra. We recall that the maximal subalgebras  of $\mathsf{AC}(1,3)$ \cite{BeckersHarnadPerroudWinternitz1978} and of the de Sitter algebra $\ort(3,2)$ \cite{PateraSharpWinternitzEtAl1977} are known.   A later work on the classification of subalgebras of the conformal algebra $\mathsf{AC}(1,n)$  can be found in
\cite{BarannykBasarabHorwathFushchych1998}. See \cite{BarannikYurik1998} for a complete classification of maximal subalgebras of rank $n$ of the conformal algebra $\mathsf{AC}(1,n)$. We note the isomorphy of the algebras $\ort(3,2)$ and $\mathsf{AC}(1,2)$ \cite{BarannykBasarabHorwathFushchych1998}. A complete group classification of the class \eqref{waveqn1} with the constraint $F_{u_t}=0$ in $(1+1)$-dimension was performed in \cite{LahnoZhdanov2005}. An enhanced complete classification of \eqref{waveqn2} in the light-cone (characteristic) coordinates using more rigorous methods has recently been carried out in \cite{BoykoLokaziukPopovych2021}. The complete group classification problem for the class of multidimensional nonlinear wave equations of the form $$u_{tt}=\nabla\cdot(f(u)\nabla u)+g(u)$$ was solved in\cite{VasilenkoYehorchenko2001}. This paper for $f(u)=1$ contains the classification results of the subclass \eqref{waveqn2} when $F(x,u)=g(u)$.

For the class \eqref{waveqn1} or its subclass \eqref{waveqn2} classification problem  remains unsolved. The results of this paper and  those of the existing subalgebra classification of Poincar\'e, extended Poincar\'e (or similitude) and conformal algebra can be used to solve this problem, at least, in dimensions $n=2$ and $n=3$. The solution of this problem would take us too far beyond the scope of the present work.

\begin{example}
We consider the Klein--Gordon equation \eqref{KG}. The symmetry condition \eqref{symmcondition2} for $F(x,u)=-V(x)u$  splits into
$$\xi^{\mu}V_{\mu}+\kappa(x)V+\Box a=0, \quad \Box b+Vb=0, \quad a_{\nu}(x)=\frac{1}{2}g_{\nu\mu}\Box\xi^{\mu}.$$

Taking into account the expressions
$$\Box x^2=2(n+1), \quad \Box (c\cdot x)x^{\alpha}=2c^{\alpha}$$ we find
$$\Box \xi^{\alpha}=[4-2(n+1)]c^{\alpha}=2(1-n)c^{\alpha}$$ and
$$a_{\nu}(x)=(1-n)c_{\nu}\quad \Rightarrow \quad a(x)=(1-n)c_{\nu}x_{\nu}+\ell=(1-n)c\cdot x+\ell,$$
where $\ell$ is a constant. So, $\Box a=0$ is automatically satisfied and the classifying equation simplifies to
\begin{equation}\label{class-eq}
  \xi^{\mu}V_{\mu}+\kappa(x)V=\xi^{\mu}V_{\mu}+2\xi_0^0V=0.
\end{equation}
$\eta$ is given by
$$\eta(x,u)=(1-n)(c\cdot x)u+\ell u+b(x),$$ where $b(x)$ runs through the solution set of the original equation,  $\Box b+Vb=0$.
The presence of the constant parameter~$\ell$ and the parameter function~$b$ are related to the homogeneity and linearity of the original equation, giving rise to the symmetries
$$S=u\gen u, \quad X(b)=b(x)\gen u.$$

For $V=0$ $(n\geq 2)$, Eq. \eqref{symmcondition2} gives the linear wave equation $\Box u=0$. In this case, \eqref{class-eq} is identically satisfied.

\end{example}

\begin{example}\label{nonlin-KG}
For the special nonlinearity $F(u)=F_0u^k$, $(k\ne 0,1)$, $n\geq 2$ the first equation of \eqref{symmcondition2} splits into the following equations
$$ka+(2\xi^0_0-a)=0, \quad kb=0,  \quad \Box a=0,  \quad \Box b=0.$$
We find $b(x)=0$, $a(x)=2/(1-k)\xi^0_0=\kappa(x)/(1-k)=(1-k)^{-1}[4(c\cdot x)+2d]$.  Substituting $a(x)$ in the second equation of \eqref{symmcondition2} it follows that we must have $4/(1-k)=1-n$ or $k=(n+3)/(n-1)$ for conformal invariance. This gives the well-known nonlinear conformally invariant wave equation (see for example \cite{FushchichShtelenSerov1993}, \cite{VasilenkoYehorchenko2001})
\begin{equation}\label{conf-inv-wave}
  \Box u=F_0 u^{(n+3)/(n-1)}.
\end{equation}
Then we can write $$\eta(x,u)=a(x)u=(1-n)[(c\cdot x)+\frac{d}{2}]u=\frac{1-n}{4}\kappa(x)u$$ and $\xi^{\mu}$ are given by the Killing solution \eqref{killing-sol}. The corresponding symmetry algebra has a basis given by \eqref{basis-thm-2}. This algebra coincides with that of the linear wave equation when the infinite-dimensional subalgebra is factored out.

It can be shown that Eq. \eqref{conf-inv-wave} is the most general wave equation invariant, up to an equivalence, under the full conformal algebra \cite{VasilenkoYehorchenko2001}.

The conformal  transformations which leave this equation invariant are given by
\begin{equation}\label{conf-trans}
  \tilde{x}^{\mu}=\frac{x^{\mu}-x^2\varepsilon^{\mu}}{1-2(\varepsilon\cdot x)+(\varepsilon\cdot x)^2},  \quad \tilde{u}(\tilde{x})=(1-2(\varepsilon\cdot x)+(\varepsilon\cdot x)^2)^{(n-1)/2}u(x),
\end{equation}
where $\varepsilon^{\mu}$, $\mu=0,1,2,\ldots, n$ are the group parameters. The first transformation of \eqref{conf-trans} is the conjugation of the translational group $\exp\{{\varepsilon}^\mu P_{\mu}\}$ by the inversions $I_{\mu}(x)=x^{\mu}/x^2$ in the unit hyperboloid $x^2=1$.
\end{example}

\begin{example}
  We are  interested in the nonlinearity $F$ with derivative dependence
\begin{equation}\label{NLW-ext}
  \Box u+f(x_0)u_{0}+g(u)=0.
\end{equation}
Obviously, the added nonlinearity destroys the full conformal symmetry of the linear wave equation $\Box u=0$. For arbitrary $f$ and $g$, this PDE is invariant under the Euclidean algebra $\euc(n)$ of dimension $n(n+1)/2$ spanned by
\begin{equation}\label{Euclidean}
  P_\mu=\partial_{\mu}, \quad J_{\mu\nu}=x_{\mu}\partial_{\nu}-x_{\nu}\partial_{\mu},  \quad \mu,\nu=1,2,\ldots,n.
\end{equation}
When $m=0$ (no dependence on derivative), for arbitrary nonlinearity $g$, the coresponding nonlinear Klein--Gordon PDE admits a Poincar\'{e} algebra $\poinc(1,n)$ as symmetry algebra. Further invariance of this equation places the conditions
\begin{equation}\label{KG-dil}
  \eta g'(u)+(2-\eta_u)g=0,  \quad \eta=au+b,
\end{equation}
where $a,b$ are constants. This leads, modulo a combination of rescaling together with shift of $u$, to two possible nonlinearities with larger symmetry algebras
\begin{equation}\label{nonlin-g-1}
  g(u)=\lambda u^{k},  \quad k\ne 0,1,
\end{equation}
and
\begin{equation}\label{nonlin-g-2}
  g(u)=\lambda e^{u}.
\end{equation}
Example \ref{nonlin-KG}  discusses the maximal symmetry algebra of the first case (wave equation with power nonlinearity \eqref{nonlin-g-1}).

We turn back to the case when $m\ne 0$ and $g$ has the special form \eqref{nonlin-g-1}. Again we solve \eqref{symmcondition1} for $$F(x_0,u,u_0)=-f(x_0)u_{0}-\lambda u^{k}$$
and $\eta=d_0 u$ by splitting with respect to $u_0$. This gives us the condition  $tf'+f=0$ for  $f$.
To sum up,  we obtain the nonlinear term $F$ in the form
$$F=\frac{m}{x_0}u_0+\lambda u^{k}, \quad d_0=\frac{2}{1-k},$$ where $m$ and $\lambda$ are integration constants. The corresponding dilational symmetry generator is
$$\tilde{D}=x^{\sigma}\partial_{\sigma}+d_0u\gen u.$$ We now look at the possibility of further extension of the extended Euclidean algebra $\tilde{\euc}(n)$ so as to include conformal symmetries. For the above choice of $F$, using part of the conformal Killing solution $\xi^{\alpha}_c=2(c\cdot x)x^{\alpha}-c^{\alpha}x^2$, $\alpha=1,2,\ldots,n$ and $\eta(x,u)=q(c\cdot x)u$ in \eqref{symmcondition1} and splitting with respect to the terms  $u^k$ and  $u_{\sigma}$, $\sigma=1,2,\ldots,n$ we find the following conditions on the parameters $m$ and $k$:
\begin{equation}\label{conf-conds}
  k=\frac{n+3+m}{n-1+m},  \quad q=(1-n)(m+1), \quad m\ne 1-n.
\end{equation}
\end{example}
So  we have  constructed the $n$-dimensional nonlinear Euler--Poisson--Darboux equation
\begin{equation}\label{EPD}
  \Box u+\frac{m}{x_0}u_0+\lambda u^k=0,
\end{equation}
where $k$ is as defined in \eqref{conf-conds}.
Eq. \eqref{EPD} is invariant under the $(n+1)(n+2)/2$-dimensional conformal group $\Conf(n)$ of the Euclidean space $\mathbb{R}^{n}$.
Additionally to the generators \eqref{Euclidean} we obtain the following ones
\begin{equation}\label{additional-sym}
  \tilde{D}=D+\frac{1}{2}(1-n-m)u\gen u, \quad
K_{\mu}=2x_{\mu}D-x^2\partial_{\mu}+(1-n)(m+1)x_{\mu}u\gen u,
\end{equation}
where $\mu=1,2,\ldots,n$ and $m\ne 1-n$.
The special choice $m=0$ takes us back to the $\Conf(1,n)$ invariant nonlinear wave equation \eqref{conf-inv-wave}.

\section{Canonical forms for symmetry vector fields} So far, we have determined the structure of the symmetry vector field $Q=\xi^{\mu}(x)\partial_{\mu} + \eta(x,u)\partial_u$, by giving the structure of $\xi^{\mu}\partial_{\mu}$ and obtaining the condition $\eta_u\neq 0$. We know that $X=\xi^{\mu}\partial_{\mu}$ must be a vector field in the Lie conformal algebra $\conf(1,n)$. The question now is to decide what form the functions $\eta$ can take, since we want to know $Z=X+\eta\partial_u$ for any given $X\in \conf(1,n)$.

A first step in answering this question is to realise that we want the vector fields $Z_{i}=X_{i}+\eta_{i}\partial_u$  to form a Lie algebra. Thus, there must be structure constants $c^k_{ij}$ so that
\begin{equation}\label{Lie-alg}
[Z_i, Z_j]=c^k_{ij}Z_k.
\end{equation}
Now, we note that $[X_i, \partial_u]=0$, so that we have, after an elementary calculation
\[
[Z_i, Z_j]=[X_i, X_j]+ [X_i(\eta_j)-X_j(\eta_i) + \eta_i\eta'_j-\eta_j\eta'_i]\partial_u,
\]
where $\eta'_i=\partial_u\eta_i$, and since we must have \eqref{Lie-alg} we then have
\[
[X_i, X_j]+ [X_i(\eta_j)-X_j(\eta_i) + \eta_i\eta'_j-\eta_j\eta'_i]\partial_u=c^k_{ij}X_k+c^k_{ij}\eta_k\partial_u.
\]
Clearly, since the $X_i$'s form the conformal Lie algebra  $\conf(1,n)$ then the structure constants $c^k_{ij}$ must be those of the conformal Lie algebra $\conf(1,n)$. With this in mind, we have the following results:
\begin{thm}\label{conformalsymmetry1} If $Z=\xi^{\mu}(x)\partial_{\mu} + \eta(x,u)\partial_u\in \conf(1,n)$, with $\xi^{\mu}(x)\partial_{\mu}\neq 0$,  is a symmetry of the equation (\ref{waveqn1}) for $n\geq 2$, then, up to an equivalence transformation of equation (\ref{waveqn1}), $Z$ is an element of the following realization of the conformal Lie algebra $\conf(1,n)$ with basis
\begin{equation}\label{basis-thm-1}
\begin{split}
P_{\mu}&=\partial_{\mu},\\
J_{\mu\nu}&=x_{\mu}\partial_{\nu}-x_{\nu}\partial_{\mu},\\
D&=x^{\sigma}\partial_{\sigma} +\varepsilon u\partial_u,\\
K_{\mu}&=2x_{\mu}D-(x_{\alpha}x^{\alpha})\partial_{\mu},
\end{split}
\end{equation}
where $\varepsilon\in\{0,1\}$.

\end{thm}

\smallskip\noindent{\bf Proof:} For the generators of translation $\partial_{\mu}$ the corresponding structure constants are zero. Let
\[
P_{\mu}=\partial_{\mu} + \eta_{\mu}\partial_u
\]
denote their extensions. Then we must have for $\mu, \nu\in \{0, 1, \dots, n\}$
\[
[P_{\mu}, P_{\nu}]=0.
\]
This gives us the system of equations
\begin{equation}\label{commutationcondition}
\partial_{\mu}\eta_{\nu}-\partial_{\nu}\eta_{\mu}=\eta_{\mu}\eta'_{\nu}-
\eta_{\nu}\eta'_{\mu},\quad \mu, \nu= 0, 1, \dots, n,
\end{equation}
where $\eta'_{\mu}=\partial \eta_{\mu}/\partial u$. Define the differential forms $\omega=\eta_{\mu}dx^{\mu},\; \pi=\eta'_{\mu}dx^{\mu}$. Then these equations can be written as
\begin{align*}
d\omega&=\pi\wedge\omega +du\wedge\pi\\
&=\pi\wedge(\omega - du).
\end{align*}
Now if we write $\Omega=\omega - du$ and note that $d\Omega=d\omega$, we have the equation
\[
d\Omega=\pi\wedge\Omega,
\]
and this means that the ideal generated by the $1$-form $\Omega$ is closed, and consequently there are functions $\alpha(x,u)$ and $\phi(x,u)$ so that
\[
\Omega=\alpha(x,u)d\phi(x,u)
\]
and in particular $\eta_{\mu}=\alpha(x,u)\partial_{\mu}\phi(x,u)$. That is, there exist solutions to the system (\ref{commutationcondition}).

The equivalence group of the class (1.1) consists of the point transformations of the form
\begin{equation}\label{conf-equiv-group}
  \tilde{x}^{\mu}=X^{\mu}(x), \quad \tilde{u}=U(x,u)
\end{equation}
such that $(X^0, X^1,\dots, X^n)\in \Conf(1,n),\;\; U_u\neq 0$.  Under such a transformation, with $\tilde{x}^{\mu}=x^{\mu}$, the vector fields  $P_{\mu}=\partial_{\mu}+\eta_{\mu}(x,u)\partial_u$ are mapped to
\[
P'_{\mu}=\partial_{\tilde{\mu}} + (P_{\mu}U)\partial_{\tilde{u}},\quad \mu= 0, 1, \dots, n.
\]
Consider now the system of $n+1$ linear homogeneous first-order partial differential equations
\[
P_{\mu}U=0,\quad \mu= 0, 1, \dots, n.
\]
Because the $P_{\mu}$ are in involution, there exists a common integral $U$ by Frobenius' theorem, and we may choose one such that $U_u\neq 0$.  Thus, there is a transformation of the equivalence group of equation (\ref{waveqn1}) which maps the $P_{\mu}$ into the canonical forms $\partial_{\mu},\,\; \mu=0, 1, \dots, n$. Hence, in our symmetry algebra, we may take our generators of space-time translations as just $\partial_{\mu},\,\; \mu=0, 1, \dots, n$.

With this choice of the space-time generators of our symmetry algebra, we have the residual equivalence group as
\[
\tilde{x}^{\mu}=x^{\mu} + k^{\mu},\quad \tilde{u}=U(u),
\]
where $k^{\mu}$ are constants and $U'(u)\neq 0$.

Now consider $\tilde{D}=D+\eta(x,u)\partial_u$. We have
\[
[\partial_{\mu}, D]=\partial_{\mu},
\]
so we must have  $[\partial_{\mu}, \tilde{D}]=\partial_{\mu}$ giving $\partial_{\mu}\eta(x,u)=0$, so that $\eta=\eta(u)$. Under a transformation $\tilde{x}^{\mu}=x^{\mu},\; \tilde{u}=U(u)$ with $U'(u)\neq 0$, $\tilde{D}$ is mapped to
\[
\tilde{D}'=D+\eta(u)U'(u)\partial_{\tilde{u}}.
\]
If $\eta(u)=0$ then we take $D=x^{\sigma}\partial_{\sigma}$ in canonical form. If $\eta(u)\neq 0$ then we take $U(u)$ so that $\eta(u)U'(u)=U(u)$ so that we have
\[
\tilde{D}=D+u\partial_u
\]
in canonical form. So we have two possible canonical forms for the dilatation operator:  $D=x^{\sigma}\partial_{\sigma}$ or $D=x^{\sigma}\partial_{\sigma} + u\partial_u$.

Now we come to the extension of the generators $J_{\mu\nu}$. Put $\tilde{J}_{\mu\nu}=J_{\mu\nu}+\eta_{\mu\nu}(x,u)\partial_u$. Since we have $$[\partial_{\lambda}, J_{\mu\nu}]=g_{\lambda\mu}\partial_{\nu}-g_{\lambda\nu}\partial_{\mu}$$
we must have $\partial_{\lambda}\eta_{\mu\nu}=0$ so that $\eta_{\mu\nu}=\eta_{\mu\nu}(u)$. Thus we have
\[
\tilde{J}_{\mu\nu}=J_{\mu\nu}+\eta_{\mu\nu}(u)\partial_u.
\]
Further, the vector fields $\eta_{\mu\nu}(u)\partial_u$ must satisfy the same commutation relations as the $J_{\mu\nu}$, so they must be a rank-one realization of the Lie algebra $\So(1,n)$ (a collection of vector fields $X_1,\dots, X_m$ is of rank one if $X_i\wedge X_j=0$ for all $i,j=1,\dots, m$). This is impossible for $n\geq 3$, since this would mean that there would be a rank-one realization of the compact Lie algebra ${\So(n)}$ for $n\geq 3$. Now each such real Lie algebra contains a subalgebra isomorphic to $\So(3)$ so we would have a rank-one realization of ${\So}(3)$. If we have the basis $Q_1, Q_2, Q_3$ of ${\So}(3)$ with $[Q_1, Q_2]=Q_3,\; [Q_2, Q_3]=Q_1,\; [Q_3, Q_1]=Q_2$, then a rank-one realization means that we may write $Q_1=fQ_3,\; Q_2=gQ_3$ for some real functions $f, g$. The commutation relations then give $fg'-gf'=1$, $f=-g'$, $g=f'$ where $f'=Q_3f$, $g'=Q_3g$. Consequently we find that $f^2+g^2=-1$ which is a contradiction for real functions $f, g$. Hence we have $\eta_{\mu\nu}=0$ for $n\geq 3$.

For $n\geq 3$, let $\tilde{K}_{\mu}=K_{\mu} + a_{\mu}(x,u)\partial_u$ be the extension of $K_{\mu}$. The commutation relations
\[
[\partial_{\mu}, K_{\nu}]=2g_{\mu\nu}D - 2J_{\mu\nu}
\]
give us, for $\partial_{\mu}a_{\nu}(x,u)=0$ so that $\tilde{K}_{\mu}=K_{\mu} + a_{\mu}(u)\partial_u$. Then, since we must have
\[
[J_{\mu\nu}, \tilde{K}_{\lambda}]=g_{\lambda\nu}\tilde{K}_{\mu}-g_{\lambda\mu}\tilde{K}_{\nu},
\]
we obtain $g_{\lambda\nu}a_{\mu}(u)-g_{\lambda\mu}a_{\nu}(u)=0$ for any choice of $\lambda, \mu, \nu$, so we conclude that $a_{\mu}(u)=0$, so that $\tilde{K}_{\mu}=K_{\mu}$.

For $n=2$, we take $\tilde{J}_{\mu\nu}=J_{\mu\nu}+\eta_{\mu\nu}(x,u)\partial_u$. From $[\partial_{\lambda}, J_{\mu\nu}]=g_{\lambda\mu}\partial_{\nu} - g_{\lambda\nu}\partial_{\mu}$, we infer that $[\partial_{\lambda}, \tilde{J}_{\mu\nu}]=g_{\lambda\mu}\partial_{\nu} - g_{\lambda\nu}\partial_{\mu}$, and this gives $\partial_{\lambda}\eta_{\mu\nu}=0$ so we have $\tilde{J}_{\mu\nu}=J_{\mu\nu}+\eta_{\mu\nu}(u)\partial_u$ and then $\eta_{\mu\nu}\partial_u$ must be a rank-one realization of ${\sf so}(1, 2)$. Since ${\sf so}(1,2)$ is a simple Lie algebra, we have either all $\eta_{\mu\nu}(u)=0$ or all $\eta_{\mu\nu}(u)\neq 0$. Assume then that all $\eta_{\mu\nu}(u)\neq 0$.

We consider $\tilde{J}_{12}=J_{12} + \eta_{12}\partial_u$. The residual equivalence group is $\tilde{x}^{\mu}=x^{\mu},\; \tilde{u}=U(u)$ with $U'(u)\neq 0$. Under such a transformation, $\tilde{J}_{12}$ is mapped to $J_{12} + \eta_{12}(u)U'(u)\partial_{\tilde{u}}$ and we choose $U(u)$ so that $\eta_{12}(u)U'(u)=1$, giving $\tilde{J}_{12}=J_{12} + \partial_u$ in canonical form. From the relations $[\tilde{J}_{12}, \tilde{J}_{01}]= \tilde{J}_{02},\; [\tilde{J}_{12}, \tilde{J}_{02}]= -\tilde{J}_{01},\; [\tilde{J}_{01}, \tilde{J}_{02}]=-\tilde{J}_{12}$ we obtain
\[
\eta_{01}'(u)=\eta_{02}(u),\quad \eta_{02}'(u)=-\eta_{01}(u),\quad \eta_{02}(u)\eta_{01}'(u)-\eta_{01}(u)\eta_{02}'(u)=1.
\]
From these we have $\eta_{01}^2(u) + \eta_{02}^2(u)=1$, so that we may write
\[
\eta_{01}(u)=\sin\phi(u),\quad \eta_{02}(u)=\cos\phi(u)
\]
for some function $\phi(u)$. Then $\eta_{02}(u)\eta_{01}'(u)-\eta_{01}(u)\eta_{02}'(u)=1$ gives $\phi'(u)=1$ so that $\phi(u)=u + k$ for some constant $k$, and then $\tilde{J}_{12}=J_{12} + \partial_u,\; \tilde{J}_{01}=J_{01} + \sin(u+k)\partial_u,\; \tilde{J}_{02}=J_{02} + \cos(u+k)\partial_u$. The residual equivalence group is $\tilde{x}^{\mu}=x^{\mu},\; \tilde{u}=u + l$, and under such a transformation  $\tilde{J}_{12}$,  $\tilde{J}_{01}$,  $\tilde{J}_{02}$ are mapped to $\tilde{J}_{12}=J_{12} + \partial_{\tilde{u}},\; \tilde{J}_{01}=J_{01} + \sin(\tilde{u}+k-l)\partial_{\tilde{u}},\; \tilde{J}_{02}=J_{02} + \cos(\tilde{u}+k-l)\partial_{\tilde{u}}$. Choosing $l=k$ we find that we have $\tilde{J}_{12}=J_{12} + \partial_u,\; \tilde{J}_{01}=J_{01} + \sin(u)\partial_u,\; \tilde{J}_{02}=J_{02} + \cos(u)\partial_u$ in canonical form.

Now consider the extension $\tilde{K}_{\mu}=K_{\mu} + a_{\mu}(x,u)\partial_u$ of $K_{\mu}$. We have the commutation relations
\[
[\partial_{\mu}, K_{\nu}]=2g_{\mu\nu}D-2J_{\mu\nu},
\]
and so we must have
\[
[\partial_{\mu}, \tilde{K}_{\nu}]=2g_{\mu\nu}D-2\tilde{J}_{\mu\nu},
\]
from which we deduce that $\partial_{\mu}a_{\nu}(x,u)=-2\eta_{\mu\nu}(u)$, and then we have $a_{\mu}(x,u)=2x^{\nu}\eta_{\mu\nu}(u)+b_{\mu}(u)$. Then we use the relations $[D, K_{\mu}]=K_{\mu}$ so that we will have $[D, \tilde{K}_{\mu}]=\tilde{K}_{\mu}$, and this gives $b_{\mu}=0$ and thus $a_{\mu}=2x^{\nu}\eta_{\mu\nu}(u)$. So $\tilde{K}_{\mu}=K_{\mu} + 2x^{\lambda}\eta_{\mu\lambda}(u)\partial_u$. We have $[K_{\mu}, K_{\nu}]=0$ so that $[\tilde{K}_{\mu}, \tilde{K}_{\nu}]=0$, and this gives
\[
K_{\mu}(x^{\lambda}\eta_{\nu\lambda}(u))- K_{\nu}(x^{\lambda}\eta_{\mu\lambda}(u)) + 2x^{\rho}x^{\sigma}[\eta_{\mu\rho}(u)\eta'_{\nu\sigma}(u) - \eta_{\nu\sigma}(u)\eta'_{\mu\rho}(u)]=0.
\]
For $n\geq 3$ we have $\eta_{\mu\nu}(u)=0$ so that $\tilde{K}_{\mu}=K_{\mu}$. For $n=2$ we may choose $\eta_{12}(u)=1,\; \eta_{01}(u)=\sin u,\; \eta_{02}(u)=\cos u$. Then considering $[\tilde{K}_0, \tilde{K}_1]$ we find that $[\tilde{K}_0, \tilde{K}_1]\neq 0$. Hence we have a contradiction, and so we must have $\eta_{\mu\nu}(u)=0$ for $n=2$.

Finally, we come to the second canonical form for $D$, namely $\tilde{D}=x^{\sigma}\partial_{\sigma} + u\partial_u$. We have, as above, $\tilde{J}_{\mu\nu}=J_{\mu\nu} + \eta_{\mu\nu}\partial_u$. As we have seen, $\eta_{\mu\nu}(u)=0$ for $n\geq 3$. We put $\tilde{K}_{\mu}=K_{\mu} + a_{\mu}(x,u)\partial_u$ as before, and then
\[
[\partial_{\mu}, \tilde{K}_{\nu}]=2g_{\mu\nu}\tilde{D}-2\tilde{J}_{\mu\nu},
\]
gives $\partial_{\mu}a_{\nu}(x,u)=2g_{\mu\nu}u - 2\eta_{\mu\nu}(u)$ so that $a_{\mu}(x,u)=2x_{\mu}u + 2 x^{\lambda}\eta_{\mu\lambda}(u) + b_{\mu}(u)$. Then we invoke $[\tilde{D}, \tilde{K}_{\mu}]=\tilde{K}_{\mu}$ to obtain
\[
2x^{\lambda}u\eta'_{\mu\lambda}(u) + ub'_{\mu}=2x^{\lambda}\eta_{\mu\lambda}(u) + b_{\mu},
\]
so that we have $u\eta'_{\mu\lambda}(u)=\eta_{\mu\lambda}(u)$ and $ub'_{\mu}=2b_{\mu}$. For $n=2$ we may choose $\eta_{12}(u)=1$ if $\eta_{\mu\nu}(u)\neq 0$. We obtain the contradiction $1=0$, so we have $\eta_{\mu\nu}(u)=0$ also for the case $n=2$. The equation $ub'_{\mu}=2b_{\mu}$ gives $b_{\mu}=c_{\mu}u^2$ for some constants $c_{\mu}$. Thus, $\tilde{K}_{\mu}=K_{\mu} +
(2x_ {\mu}u + c_{\mu}u^2)\partial_u$, which we may write as $\tilde{K}_{\mu}=2x_{\mu}\tilde{D} -(x_{\lambda}x^{\lambda})\partial_{\mu} + c_{\mu}u^2\partial_u$.  Then the requirement $[\tilde{K}_{\mu}, \tilde{K}_{\nu}]=0$ gives $x_{\mu}c_{\nu} - x_{\nu}c_{\mu}=0$ for all $\mu, \nu$, and these equations can be satisfied only if $c_{\mu}=0$ for all $\mu$. Hence we have
\[
\tilde{K}_{\mu}=2x_{\mu}\tilde{D} -(x_{\lambda}x^{\lambda})\partial_{\mu},\quad \mu= 0,1, \dots, n
\]
for all $n\geq 2$. Thus we have two possible canonical forms  \eqref{basis-thm-1}.

\begin{thm}\label{conformalsymmetry2} If $Z=\xi^{\mu}(x)\partial_{\mu} + [a(x)u+b(x)]\partial_u\in \conf(1,n)$, with $\xi^{\mu}(x)\partial_{\mu}\neq 0$,  is a symmetry of the equation (\ref{waveqn2}) for $n\geq 2$, then, up to an equivalence transformation of equation (\ref{waveqn2}), $Z$ is an element of the following realization of the conformal Lie algebra $\conf(1,n)$ with basis
\begin{equation}\label{basis-thm-2}
\begin{split}
P_{\mu}&=\partial_{\mu},\\
J_{\mu\nu}&=x_{\mu}\partial_{\nu}-x_{\nu}\partial_{\mu},\\
D&=x^{\sigma}\partial_{\sigma} + \frac{(1-n)}{2}u\partial_u,\\
K_{\mu}&=2x_{\mu}D-(x_{\alpha}x^{\alpha})\partial_{\mu}+(1-n)x_{\mu}u\gen u.
\end{split}
\end{equation}
\end{thm}

\smallskip\noindent{\bf Proof:} As in the previous proof, we begin by looking at the extensions
\[
P_{\mu}= \partial_{\mu} + [a_{\mu}(x)u+b_{\mu}(x)]\partial_u.
\]
From equation (\ref{symmcondition2}) it follows that $a_{\mu}(x)$ are constants, which we denote by $a_{\mu}$.  Commutativity $[P_{\mu}, P_{\nu}]=0$ gives us the system of equations
\[
\partial_{\mu}b_{\nu}-\partial_{\nu}b_{\mu}=a_{\mu}b_{\nu}-a_{\nu}b_{\mu}.
\]
On putting
$$\omega=b_{\mu}(x)dx^{\mu},\quad \pi=a_{\mu}dx^{\mu}=d\alpha$$ with $\alpha=a_{\mu}x^{\mu}$, our system of equations can be written as
\[
d\omega=d\alpha\wedge \omega
\]
and we easily see that this can be rewritten as $d\left(e^{-\alpha}\omega\right)=0$, so there exists (locally) a function $p(x)$ so that $e^{-\alpha}\omega=dp(x)$. Hence $\omega=e^{\alpha}dp(x)$, so there are solutions to our system of equations.

Under an equivalence transformation $\tilde{x}^{\mu}=x^{\mu},\; \tilde{u}=Au+B(x)$ with $A\neq 0$ (by Proposition \ref{equivalence2} we have an equivalence transformation
$$\tilde{x}^{\mu}=X^{\mu},\quad \tilde{u}=A(x)u + B(x)$$ with $\Box X^{\sigma}=2g^{\mu\nu}X^{\sigma}_{\mu}A_{\nu}/A$, so that in our case we obtain $2g^{\mu\nu}g^{\sigma}_{\mu}A_{\nu}=0$, which gives $A_{\nu}=0$ and so $A(x)=A$, a constant) $P_{\mu}$ is transformed to
\[
P'_{\mu}=\partial_{\mu} + [a_{\mu}\tilde{u} + \partial_{\mu}B(x)-a_{\mu}B(x)+ Ab_{\mu}(x)]\partial_{\tilde{u}}.
\]
We can choose $B(x)$ so that
$$\partial_{\mu}B(x)-a_{\mu}B(x)+ Ab_{\mu}(x)=0.$$
In fact, with $\Omega=[\partial_{\mu}B(x)-a_{\mu}B(x)+ Ab_{\mu}(x)]dx^{\mu}$ we have
\begin{align*}
\Omega&=dB(x)-B(x)d\alpha + A\omega\\
&=dB(x)-B(x)d\alpha + Ae^{\alpha}dp(x)\\
&=e^{\alpha}\left[e^{-\alpha}(dB(x)-B(x)d\alpha) + Adp(x)\right]\\
&=e^{\alpha}\left[d\left(e^{-\alpha}B(x)\right) + Adp(x)\right]\\
&=e^{\alpha}d\left(e^{-\alpha}B(x) + Ap(x)\right).
\end{align*}
We see that we may always choose $B(x)$ so that $e^{-\alpha}B(x) + Ap(x)=0$ so we obtain
\[
P_{\mu}=\partial_{\mu} + a_{\mu}u\partial_u
\]
in canonical form.

Next we come to the Lorentz rotations $J_{\mu\nu}$. Their extensions must be of the form
\[
\tilde{J}_{\mu\nu}=J_{\mu\nu} + [a_{\mu\nu}u+b_{\mu\nu}(x)]\partial_u,
\]
where the $a_{\mu\nu}$ are constants, by equation (\ref{symmcondition2}). The commutation relations
\[
[P_{\lambda}, J_{\mu\nu}]=g_{\lambda\mu}P_{\nu} - g_{\lambda\nu}P_{\mu}
\]
give us the system of equations
\[
g_{\lambda\nu}a_{\mu}-g_{\lambda\mu}a_{\nu}=0,\quad \partial_{\lambda}b_{\mu\nu}(x)=a_{\lambda}b_{\mu\nu}(x),
\]
from which we deduce that $a_{\mu}=0$ and $b_{\mu\nu}(x)$ are constants, for $n\geq 2$. Hence we have
\[
\tilde{J}_{\mu\nu}=J_{\mu\nu} + [a_{\mu\nu}u+b_{\mu\nu}]\partial_u,
\]
where the $a_{\mu\nu}, b_{\mu\nu}$ are constants. The commutation relations for the $J_{\mu\nu}$ then give $a_{\mu\nu}=b_{\mu\nu}=0$, so that $\tilde{J}_{\mu\nu}=J_{\mu\nu}$ and $P_{\mu}=\partial_{\mu}$.

We now come to the extension
\[
\tilde{D}=D + [qu + b(x)]\partial_u
\]
of $D=x^{\sigma}\partial_{\sigma}$ where $q=$ constant, by equation (\ref{symmcondition2}). We have $[P_{\mu}, \tilde{D}]=P_{\mu}$ which gives us $b(x)=\text{const.}$ So
\[
\tilde{D}=D + [qu + b]\partial_u.
\]
Our residual equivalence group consists of the transformations $\tilde{x}^{\mu}=x^{\mu},\; \tilde{u}=Au + B$ with $A\neq 0$. Under such a transformation, $\tilde{D}$ is mapped to
\[
\tilde{D}'=D + [q\tilde{u} + Ab-qB]\partial_u.
\]
If $q\neq 0$ we may choose $B$ so that $Ab-qB=0$ giving $\tilde{D}=D + qu\partial_u$ in canonical form. If $q=0$ then for $b\neq 0$ we choose $A$ so that $Ab=1$, giving $\tilde{D}=D + \partial_u$ in canonical form. If $q=b=0$ then $\tilde{D}=D$. Consequently we have the canonical forms
\[
\tilde{D}=D + \partial_u,\quad \tilde{D}=D + qu\partial_u,\;\; q\in \mathbb{R}.
\]

Finally we come to the extensions

\[
\tilde{K}_{\mu}=K_{\mu} + [a_{\mu}(x)u + b_{\mu}(x)]\partial_u,
\]
where $K_{\mu}=2x_{\mu}D-(x_{\alpha}x^{\alpha})\partial_{\mu}$ and $D=x^{\sigma}\partial_{\sigma}$. We can write $K_{\mu}$ as $K_{\mu}=\xi^{\sigma}\partial_{\sigma}$ with $\xi^{\sigma}=2x_{\mu}x^{\sigma}-(x_{\rho}x^{\rho})g^{\sigma}_{\mu}$. Then using equation (\ref{symmcondition2}), we have $\Box\xi^{\sigma}=2g^{\sigma\lambda}\partial_{\lambda}a_{\mu}(x)$ and then a straightforward calculation yields $\partial_{\lambda}a_{\mu}(x)=(1-n)g_{\lambda\mu}$ so that we have $a_{\mu}(x)=(1-n)x_{\mu} + a_{\mu}$ where the $a_{\mu}$ are now constants. Thus our allowed extensions are
\[
\tilde{K}_{\mu}=K_{\mu} + [((1-n)x_{\mu} + a_{\mu})u + b_{\mu}(x)]\partial_u.
\]

We require $[P_{\lambda}, \tilde{K}_{\mu}]=2(g_{\lambda\mu}\tilde{D}-J_{\lambda\mu})$. For $\tilde{D}=D + \partial_u$ this gives
\[
[P_{\lambda}, \tilde{K}_{\mu}]=2(g_{\lambda\mu}\tilde{D}-J_{\lambda\mu}) + [(1-n)g_{\lambda\mu}u +\partial_{\lambda}b_{\mu}(x)-2g_{\lambda\mu}]\partial_u,
\]
so we must have $(1-n)g_{\lambda\mu}u +\partial_{\lambda}b_{\mu}(x)-2g_{\lambda\mu}=0$, which is impossible for $n\geq 2$. Thus we cannot have $\tilde{D}=D + \partial_u$. For $\tilde{D}=D+qu\partial_u$, with $q\in \mathbb{R}$, we obtain
\[
[P_{\lambda}, \tilde{K}_{\mu}]=2(g_{\lambda\mu}\tilde{D}-J_{\lambda\mu}) + [\left((1-n)-2q\right)g_{\lambda\mu}u +\partial_{\lambda}b_{\mu}(x)]\partial_u,
\]
from which we deduce that we must have $((1-n)-2q)g_{\lambda\mu}u +\partial_{\lambda}b_{\mu}(x)=0$. Hence $\displaystyle q=\frac{(1-n)}{2}$ and $\partial_{\lambda}b_{\mu}(x)=0$. This gives
\[
\tilde{K}_{\mu}=2x_ {\mu}\tilde{D}-(x_{\rho}x^{\rho})\partial_{\mu} + b_{\mu}\partial_u,
\]
where we now have $\tilde{D}=D+\frac{(1-n)}{2}u\partial_u$.

The commutation relations
\[
[\tilde{K}_{\lambda}, J_{\mu\nu}]=g_{\lambda\mu}\tilde{K}_{\nu} - g_{\lambda\nu}\tilde{K}_{\mu}
\]
give us the system $g_{\lambda\mu}b_{\nu} - g_{\lambda\nu}b_{\mu}=0$ so that we have $b_{\mu}=0$ for $n\geq 2$. This establishes the result.

\section{Symmetries of the form $\eta(x,u)\partial_u$} We now look at symmetries of the form $Q=\eta(x,u)\partial_u$. It is obvious that such symmetries can occur: for instance, some equations from the class \eqref{waveqn1} admit $Q=\partial_u$ as a Lie symmetry vector field. Now we take a more systematic look at the types of Lie algebras that are admissible and realised by vector fields of the form $Q=\eta(x,u)\partial_u$. Our first result shows that no semi-simple Lie algebra is admissible:

\begin{prop} No semi-simple Lie algebra realised in terms of vector fields of the form $Q=\eta(x,u)\partial_u$ can be admitted as symmetries of equation (\ref{waveqn1}).
\end{prop}

\smallskip\noindent{\bf Proof:} First, note that any Lie algebra realised by vector fields of the form $Q=\eta(x,u)\partial_u$ is a rank-one realisation.

The Lie algebra $\So(3, \mathbb{R})$ has no rank-one realisation. In fact, if we have $Q_1=\eta_1(x,u)\partial_u,\; Q_2=\eta_2(x,u)\partial_u, Q_3=\eta_3(x,u)\partial_u$ with $[Q_1, Q_2]=Q_3,\; [Q_2, Q_3]=Q_1, \; [Q_3, Q_1]=Q_2$, then we have $\eta_1\eta'_2-\eta_2 \eta'_1=\eta_3,\; \eta_2\eta'_3-\eta_3 \eta'_2=\eta_1,\; \eta_3\eta'_1-\eta_1 \eta'_3=\eta_2$, where $\eta'=\partial_u\eta(x,u)$, and a simple calculation now gives $\eta^2_1 + \eta^2_2 + \eta^2_3=0$ so that $Q_1=Q_2=Q_3=0$.

For the Lie algebra $\Sl(2, \mathbb{R})$ we take a basis $\langle Q_1, Q_2, Q_3\rangle$ with $[Q_1, Q_2]=2Q_2,\; [Q_1, Q_3]=-2Q_3,\; [Q_2, Q_3]=Q_1$. Then note that the equivalence group of equation (\ref{waveqn1}) contains a transformation $\tilde{x}^{\mu}=x^{\mu},\; \tilde{u}=U(x,u)$ with $U_u\neq 0$ and under such a transformation, $Q_3=\eta_3(x,u)\partial_u$ is mapped to $Q'_3=\eta_3(x,u)U_u(x,u)\partial_{\tilde{u}}$. We choose $U(x,u)$ so that $\eta_3(x,u)U_u(x,u)=1$, giving $Q_3=\partial_u$ in canonical form. For $Q_1=\eta_1(x,u)\partial_u$ we must have, from the commutation relation $[Q_1, Q_3]=-2Q_3$, that $\eta'_1=2$, and so $Q_1=(2u + a(x))\partial_u$. The residual equivalence group leaving $Q_3=\partial_u$ invariant in form is now $\tilde{x}^{\mu}=x^{\mu},\; \tilde{u}=u + k(x)$, and under such a transformation $Q_1$ is mapped to $$Q'_1=(2u+a(x))\partial_{\tilde{u}}=(2\tilde{u}+a(x) -2 k(x))\partial_{\tilde{u}}$$ and on choosing $k(x)$ so that $a(x)-2k(x)=0$ we obtain $Q_1=2u\partial_u$ in canonical form. Then, putting $Q_2=\eta_2(x,u)\partial_u$, the commutation relation $[Q_2, Q_3]=Q_1$ gives $\eta'_2=-2u$ so that $\eta_2=-u^2+b(x)$. With $Q_2=(-u^2+b(x))\partial_u$, the commutation relation $[Q_1, Q_2]=2Q_2$ gives us $-2u^2-2b(x)=2(-u^2+b(x))$ so that $b(x)=0$ and we then have $Q_1=-u^2\partial_u,\; Q_2=2u\partial_u,\; Q_3=\partial_u$ in canonical form. Putting these into the symmetry condition \eqref{symmcondition1} for $F$ we obtain a contradiction. The symmetry $\partial_u$ gives $F_u=0$, then $2u\partial_u$ gives $u_{\mu}F_{u_{\mu}}=F$ and finally $-u^2\partial_u$ gives $2uu_{\mu}F_{u_{\mu}}=2uF + 2g^{\mu\nu}u_{\mu}u_{\nu}$ which is a contradiction. Thus, $\text{sl}(2, \mathbb{R})$ is not realisable as symmetries of the form $Q=\eta(x,u)\partial_u$.

Since any finite-dimensional real semi-simple Lie algebra contains a subalgebra isomorphic to $\text{so}(3, \mathbb{R})$ or $\text{sl}(2, \mathbb{R})$ (see the Appendix), it now follows that no semi-simple Lie algebra can be realised as a symmetry algebra of equation (\ref{waveqn1}) in terms of vector fields $Q=\eta(x,u)\partial_u$.

Consequently we have the following statement:

\begin{cor} Any Lie algebra of symmetries of equation (\ref{waveqn1}) realised by vector fields of the form $Q=\eta(x,u)\partial_u$ is solvable.
\end{cor}

It is clear from the above results that any given equation of the form (\ref{waveqn1}) will admit a symmetry algebra which is a subalgebra of $\conf(1,n)\uplus \mathsf{A}$, where $\mathsf{A}$ is a solvable Lie algebra.

\subsection{Preliminary ideas on abelian algebras}
For purely abelian algebras we can proceed as follows: Under the conformal equivalence group \eqref{conf-equiv-group}, any single vector field of the form $\eta(x,u)\gen u$ is mapped to $\gen{u'}$ with the choice $\eta(x,u)U_u=1$ so we can start with the one-dimensional abelian algebra $\gen u$.

Now if $\langle\gen u, \eta(x,u)\gen u\rangle$ is abelian, then we have $\eta=\eta(x)$. If the vector $\nabla \eta=(\partial_0\eta, \partial_1\eta, \ldots, \partial_n\eta)$ is timelike, we may choose $\eta(x)=X^0(x)\equiv T(x)$, reducing $\langle\gen u, t\gen u \rangle$ to canonical form. If $\nabla \eta$ is spacelike we may choose $\eta(x)=X^1$ (or any other $X^a$, $a=2,\ldots, n$); if $\nabla \eta$ is timelike, then we may choose $\eta=T-X^1$. Thus we have the following canonical forms:
$$\langle\gen u, t\gen u\rangle,  \quad \langle\gen u, x_1\gen u\rangle, \quad  \langle\gen u, (t+x_1)\gen u\rangle.$$ We note that $x_1=-x^1$ because of the Lorentzian metric.

For three-dimensional abelian algebras we can proceed as follows. If we have $\langle\gen u, t\gen u, \eta(x)\gen u\rangle$ then $\partial_a \eta\not =0$ for some $a=1,2,\ldots,n$; for otherwise we would have an algebra $\langle\gen u, t\gen u, \eta(t)\gen u\rangle$ and with $\eta''(t)\ne 0$ because of linear independence; however this leads to a contradiction. Note that any permutation of  $X^1, X^2, \ldots, X^n$ is an equivalence transformation, so we can assume $\partial_1 \eta\ne 0$. Then $\nabla \eta$ is spacelike, and we can choose $X^1$ so that $\nabla \eta=\nabla X^1$, so that $\nabla (\eta-X^1)=0$, giving $\eta=X^1+g(t)$, and then we find that we have $\langle\gen u, t\gen u, (x_1+g(t))\gen u\rangle$ in canonical form. This algebra gives $F$ being independent of $u_0$ but linear in $u$: $F=g''(t)u_1+\tilde{F}(x,u_2,\ldots,u_n)$.

For the two-dimensional nonabelian algebra we have the canonical form $\langle\gen u,u\gen u\rangle$. In this case, $F$ satisfies
$$F_u=0, \quad u_{\mu}F_{u_\mu}-F=0,$$ leading to
$$F=u_0\tilde{F}(x,\omega_1,\ldots,\omega_n),  \quad \omega_a=u_0^{-1}u_{a}, \quad a=1,2,\ldots,n.$$

A complete classification of all solvable algebras remains open.

\section{Appendix}

\begin{lemma} If ${\sf g}$ is a real, simple Lie algebra, then it contains a subalgebra isomorphic to $\So(3, \mathbb{R})$ if ${\sf g}$ is compact or it contains a subalgebra isomorphic to $\Sl(2, \mathbb{R})$ if ${\sf g}$ is non-compact.
\end{lemma}

\smallskip\noindent{\bf Proof:} First, assume that ${\sf g}$ is compact. Then it is a compact real form of the complexification ${\sf g}_\mathbb{C}={\sf g}\oplus i{\sf g}$ which is simple. Any two compact real forms of a semi-simple complex Lie algebra are conjugate to each other in ${\sf g}_\mathbb{C}.$ The complex Lie algebra ${\sf g}_\mathbb{C}$ has a root-space decomposition
\[
{\sf g}= {\sf h} + \bigoplus_{\alpha\in \triangle}{\sf g}_{\alpha},
\]
where ${\sf h}$ is a Cartan subalgebra of ${\sf g}_\mathbb{C}.$ The root spaces ${\sf g}_{\alpha}$ are one-dimensional and ${\sf g}_{-\alpha}$ is a root space if ${\sf g}_{\alpha}$ is a root space.  We denote the basis vectors of the ${\sf g}_{\alpha}$ by  $e_{\alpha}$  and they act nilpotently on ${\sf g}.$  For each root $\alpha$ there is an element $h_{\alpha}\in {\sf h}$ so that
\[
[e_{\alpha}, e_{-\alpha}]=h_{\alpha}, \quad [h_{\alpha}, e_{\alpha}]=2e_{\alpha}, \quad [h_{\alpha}, e_{-\alpha}]=-2e_{-\alpha}.
\]
Then the real subspace of ${\sf g}_\mathbb{C}$ defined by
\[
\bigoplus_{\alpha\in \triangle}\mathbb{R}i{\sf h}_{\alpha}\bigoplus_{\alpha\in \triangle}\mathbb{R}(e_{\alpha}-e_{-\alpha}) \bigoplus_{\alpha\in \triangle} \mathbb{R}i(e_{\alpha}+e_{-\alpha})
\]
is a compact real form of ${\sf g}_\mathbb{C}.$ Then define for some $\alpha$
\[
f=\frac{1}{2}(e_{\alpha}-e_{-\alpha}),\quad g=\frac{i}{2}( e_{\alpha}+e_{-\alpha}), \quad h=\frac{i}{2}h_{\alpha}
\]
and we have
\[
[f,g]=h,\quad [g,h]=f, \quad [h,f]=g
\]
which are the commutation relations for $\So(3, \mathbb{R}).$ This real Lie algebra is conjugate to ${\sf g}$ and therefore ${\sf g}$ must contain a subalgebra which is isomorphic to $\So(3, \mathbb{R}).$

Next, if ${\sf g}$ is non-compact, then ${\sf g}$ has a non-trivial Iwasawa decomposition
\[
{\sf g}={\sf k} + {\sf a} + {\sf n},
\]
where ${\sf a}$ is abelian, ${\sf n}\neq \{0\}$ is nilpotent and acts nilpotently on ${\sf g}.$ The Jacobson-Morozow theorem states that a nilpotent element in a semi-simple Lie algebra over a field of characteristic zero (in our case, $\mathbb{R}$) can be embedded in a subalgebra isomorphic to $\Sl(2, \mathbb{R}).$ This establishes the result.



\begin{thebibliography}{10}

\bibitem{Bateman1909}
H.~Bateman.
\newblock {T}he conformal transformations of a space of four dimensions and
  their applications to geometrical optics.
\newblock {\em Proc. London Math. Soc.}, 7:70--89, 1909.

\bibitem{Cunningham1910}
E.~Cunningham.
\newblock {The principle of relativity in electrodynamics and an extension
  thereof}.
\newblock {\em Proc. London Math. Soc.}, s2-8(1):77--98,
  01 1910.

\bibitem{Carmichael1927}
R.~D. Carmichael.
\newblock Transformations leaving invariant certain partial differential
  equations of physics.
\newblock {\em American J. of Math.}, 49(1):97--116, 1927.

\bibitem{CrampinPirani1987}
M.~Crampin and F.~A.~E. Pirani.
\newblock {\em Applicable Differential Geometry}.
\newblock Cambridge University Press, Cambridge, 1987.

\bibitem{Schottenloher2008}
M. Schottenloher.
\newblock {\em A Mathematical Introduction to Conformal Field Theory}.
\newblock Lecture Notes in Physics. Springer-Verlag Berlin Heidelberg, 2008.

\bibitem{FushchichShtelenSerov1993}
W.~I. Fushchich, W.~M. Shtelen, and N.~I. Serov.
\newblock {\em {S}ymmetry Analysis and Exact Solutions of Equations of
  Nonlinear Mathematical Physics}.
\newblock Kluwer Academic Publishers, Dordrecht, 1993.

\bibitem{BeckersHarnadPerroudWinternitz1978}
J.~Beckers, J.~Harnad, M.~Perroud, and P.~Winternitz.
\newblock Tensor fields invariant under subgroups of the conformal group of
  space-time.
\newblock {\em J. Math. Phys.}, 19(10):2126--2153, 1978.

\bibitem{PateraSharpWinternitzEtAl1977}
J.~Patera, R. T. Sharp, P.~Winternitz, and H.~Zassenhaus.
\newblock {Continuous} {subgroups} of the {fundamental} {groups} of {physics}.
  {III} {the} de {Sitter} {groups}.
\newblock {\em J. Math. Phys}, 18(12):2259--2288, 1977.

\bibitem{BarannykBasarabHorwathFushchych1998}
L.~F. Barannyk, P.~Basarab-Horwath, and W.~I. Fushchych.
\newblock On the classification of subalgebras of the conformal algebra with
  respect to inner automorphisms.
\newblock {\em J. Math. Phys.}, 39(9):4899--4922, 1998.

\bibitem{BarannikYurik1998}
A.~F. Barannik and I.~I. Yurik.
\newblock Classification of maximal subalgebras of rank $n$ of the conformal
  algebra $\mathsf{AC}(1, n)$.
\newblock {\em Ukrainian Math. J.}, 50(4):519--532, 1998.

\bibitem{LahnoZhdanov2005}
V.~Lahno and R.~Zhdanov.
\newblock {G}roup classification of nonlinear wave equations.
\newblock {\em J. Math. Phys.}, 46:053301, 2005.

\bibitem{BoykoLokaziukPopovych2021}
V.~M. Boyko, O.~V. Lokaziuk, and R.~O. Popovych.
\newblock Realizations of Lie algebras on the line and the new group
  classification of (1+1)-dimensional generalized nonlinear Klein--Gordon
  equations.
\newblock {\em Anal. and Math. Phys.}, 11(3):127, 2021.

\bibitem{VasilenkoYehorchenko2001}
O.~F. Vasilenko and I.~A. Yehorchenko.
\newblock Group classification of multidimensional nonlinear wave equations.
\newblock {\em Proceedings of Institute of Mathematics of NAS of Ukraine},
  pages 63--66, 2001
\newblock (in Ukrainian).


\end{thebibliography}

\end{document}